\documentclass[aps,prb,twocolumn,amsfonts,amsmath,amssymb,superscriptaddress,showpacs,]{revtex4-1}

\usepackage{graphicx}
\usepackage[utf8]{inputenc}
\usepackage{subfigure}
\usepackage{listings}
\usepackage{bm}
\usepackage{setspace}
\usepackage{hyperref}
\usepackage{natbib}
\usepackage[normalem]{ulem}
\usepackage{epstopdf}
\usepackage{subfigure}
\usepackage{color}
\usepackage{dsfont}
\usepackage[bbgreekl]{mathbbol}
\usepackage{amsfonts}

\def\e{\mathrm{e}}
\def\d{\mathrm{d}}
\def\i{\mathrm{i}}

\makeatother
\begin{document}

\title{Spectral properties and phase diagram of correlated lattice bosons in an optical cavity within the B-DMFT}

\author{Jaromir Panas}
\affiliation{ Institute of Theoretical Physics, Faculty of Physics, University of Warsaw, Pasteura 5, 02-093 Warszawa, Poland }

\author{Anna Kauch}
\affiliation{ Institute of Physics, Academy of Sciences of the Czech Republic, Na Slovance 2, 18221 Praha, Czech Republic }
\affiliation{ Institute of Solid State Physics, TU Wien, Wiedner-Hauptstrasse 8-10/E138, 1040 Wien, Austria }

\author{Krzysztof Byczuk}
\affiliation{ Institute of Theoretical Physics, Faculty of Physics, University of Warsaw, Pasteura 5, 02-093 Warszawa, Poland }

\date{\today}

\begin{abstract}
We use the Bose-Hubbard model with an effective infinite-range interaction to describe the correlated lattice bosons in an optical cavity. We study both static and spectral properties of such system within the bosonic dynamical mean-field theory (B-DMFT), which is the state of the art method for strongly correlated bosonic systems. Both similarities and differences are found and discussed between our results and these obtained within different theoretical methods and experiment.
\end{abstract}


\maketitle


\section{Introduction}

The development of experiments with cold quantum gases in optical lattices\cite{bloch_short,bloch_long} led to a breakthrough in the studies of strongly correlated systems. Its close correspondence with the Bose-Hubbard model, together with the possibility of fine-tuning of the parameters of the system gives a remarkably powerful tool for investigating quantum phenomena in this model.\cite{jaksch} On the other hand, the fast growing field of research on cold atoms in cavity-generated optical potential gives us a good understanding of processes in which atoms interact with radiation field.\cite{RMP_cavity} Combining these two fields of research together opens up a possibility of a new fascinating study. Putting an optical lattice inside an optical cavity results in an effective infinite-range interaction between particles in the system.\cite{Landig_cavity} This long-range interaction, mediated by the cavity mode of the light, competes with the inherent short range interaction of the Bose-Hubbard model. As a result of this competition between correlations on different length scales, new states of matter emerge. On top of the phases of the Mott insulator (MI) and superfluid (SF), known from theoretical predictions\cite{fisher} and confirmed in experiment\cite{bloch_short}, we expect new phases of density wave (DW) and supersolid (SS).\cite{hof_cavity}

Recent experiment with lattice bosons in an optical cavity\cite{Landig_cavity, reza_exp} has stimulated a lot of theoretical research on this subject. In several published papers, the Bose-Hubbard model with infinite-range interaction has been studied within the static mean-field theory.\cite{Chen_mf_cavity,Dogra_cavity} These studies include results for the phase diagram of such a system and some initial results for low energy spectra. However, these mean-field type approaches treat the kinetic term of the Hamiltonian as a small perturbation. A more advanced way, to study the Bose-Hubbard model with infinite-range interaction would be to use the bosonic dynamical mean-field theory (B-DMFT).\cite{byczuk_bdmft} Such an approach allows us to obtain reliable results for any ratio of the kinetic and potential energies. First application of the B-DMFT to a system with the optical lattice inside an optical cavity was presented in Ref. \onlinecite{hof_cavity}, in which DW and SS phases were obtained. Another B-DMFT study, in Ref. \onlinecite{reza_theo}, elaborates on this topic showing a phase diagram which is directly comparable with experiment of Ref. \onlinecite{reza_exp}.

In this paper we aim to expand on the previous B-DMFT studies. Instead of performing calculations in the real-space, and thus being restricted to finite size of the lattice, we consider an infinite system and derive an appropriate self-consistency condition for a bipartite lattice in two dimensions taking into account the possibility of spontaneous breaking of translational symmetry. We obtain the phase diagram in a different parameter space, in order to compare the static mean-field and B-DMFT results. Between these two approaches we observe similarities but also significant discrepancies, revealed in the behavior of the system close to the phase transition between the SS and DW phases and in the behavior of the SS phase. We also present the spectral properties of the Bose-Hubbard model with infinite-range interaction.

This paper is organized as follows. In Sec. \ref{sect2} we introduce the Bose-Hubbard model with infinite range interaction and present the B-DMFT method together with its new self-consistency relations, appropriate for a bipartite lattice (with supplementary material included in the Appendix). We discuss some issues related to using a self-consistent approach in the studied problem in Sec. \ref{sect2c}. In Sec. \ref{sect3a} we present the phase diagram of the system, compare it to the one obtained within the static mean-field approach and discuss differences between the results of the two approaches. In Sec. \ref{sect3b} we present the local densities of states and momentum resolved spectral functions and analyze their features. In Sec. \ref{sect4} we provide the summary of our results.


\section{Model and investigation method}\label{sect2}


\subsection{The Bose-Hubbard model with cavity mediated infinite-range interaction}\label{sect2a}

We consider a system with cold-atom quantum gas trapped in an optical lattice which is additionally placed inside an optical cavity. Such a setup was recently realized in experiment.\cite{reza_exp, Landig_cavity} The counter-propagating laser beams of wavelength $\lambda$ create a standing wave. This results in an effective periodic potential, which has the periodicity equal to half of the wavelength of the light beam, $\lambda/2$. We consider a two dimensional (2D) realization of such a system in the $xz$ plane. The laser beam in the $z$ direction plays a second role as it drives a cavity mode in the $x$ direction through scattering of light on atoms in the system. The scattering processes between atoms and the cavity light creates a $\lambda$-periodic modulation of the optical-lattice potential. Theoretical treatment of such experiments requires a complex analysis of a system with many degrees of freedom: (i) {\it internal atomic degrees of freedom} - processes of exciting an electron in an atom (it is justified to treat an atom as a two state system\cite{RMP_cavity}), (ii) {\it a light mode in the cavity degrees of freedom} -  processes of creating and annihilating photons in the cavity due to the scattering of photons on atoms, (iii) {\it motional degrees of freedom} - an atom moving through the system, hopping from one potential well to a neighboring one.\cite{RMP_cavity} Within dispersive limit, when atomic saturation effects are negligible and atoms are considered as linearly polarizable particles, one can get rid of atomic internal degrees of freedom.\cite{RMP_cavity} If the decay rate of photons from the cavity is large we can adiabatically eliminate the cavity field. Then we obtain an effective Hamiltonian with an infinite-range interaction mediated by the cavity mode in the following form\cite{Landig_cavity}
\begin{equation}\label{hamiltonian}
\begin{split}
\hat{H}&=-\sum_{i,j} t_{ij} \hat{b}^{\dag}_{i} \hat{b}^{\phantom\dag}_j - \mu \sum_{i} \hat{b}^{\dag}_{i} \hat{b}^{\phantom\dag}_i  + \frac{U}{2}\sum_{i} \hat{b}^{\dag}_i \hat{b}^{\dag}_i \hat{b}^{\phantom\dag}_i \hat{b}^{\phantom\dag}_i \\
&-\frac{V}{N} \left(\sum_{i\in S_A}\hat{b}^{\dag}_{i} \hat{b}^{\phantom\dag}_i -\sum_{i\in S_B}\hat{b}^{\dag}_{i} \hat{b}^{\phantom\dag}_i\right)^2,
\end{split}
\end{equation}
where $ \hat{b}^{\dag}_i $ ($ \hat{b}_i $) is a bosonic creation (annihilation) operator on a lattice site $ i $, $\mu$ is the chemical potential, $U$ is the local interaction strength, and $ t_{ij} $ is the hopping amplitude. The first three terms represent the Bose-Hubbard Hamiltonian for which we assume nearest neighbor (NN) hopping, i.e., $t_{ij}=t>0$ if sites $i$ and $j$ are NN and $t_{ij}=0$ otherwise. These terms alone describe a homogeneous isotropic square lattice and would correspond to a system without the cavity field. The last term in the Hamiltonian represents an effective infinite-range interaction, mediated by the cavity field. Such an interaction splits the square lattice into two sublattices $A$ and $B$. The parameter $V$ controls the strength of this interaction. $S_A$ and $S_B$ denote sets of site indices corresponding to the sublattices $A$ and $B$, respectively.
Because of the $N^{-1}$ term in the last part of the Hamiltonian (\ref{hamiltonian}) the fluctuations are negligible for this type of interaction in the thermodynamic limit. Therefore it is sufficient to treat the last term of (\ref{hamiltonian}) within a mean-field approach, which leads to the following Hamiltonian
\begin{equation}\label{hamiltonianMF}
\begin{split}
\hat{H}&=-\sum_{i,j} t_{ij} \hat{b}^{\dag}_{i} \hat{b}^{\phantom\dag}_j - \mu \sum_{i} \hat{b}^{\dag}_{i} \hat{b}^{\phantom\dag}_i  + \frac{U}{2}\sum_{i} \hat{b}^{\dag}_i \hat{b}^{\dag}_i \hat{b}^{\phantom\dag}_i \hat{b}^{\phantom\dag}_i \\
&-V \left(\sum_{i\in S_A}\hat{b}^{\dag}_{i} \hat{b}^{\phantom\dag}_i -\sum_{i\in S_B}\hat{b}^{\dag}_{i} \hat{b}^{\phantom\dag}_i\right)\left( n_A - n_B\right)\\
&+NV\frac{\left( n_A - n_B\right)^2}{4},
\end{split}
\end{equation}
where $n_A$ ($n_B$) is the average occupation of a site on the sublattice $A$ ($B$). The last term is important for the correct determination of the phase transition lines.

The Hamiltonian (\ref{hamiltonianMF}), is of the form of the Bose-Hubbard model with the addition of an effective staggered mean-field, resulting in a lattice with $A$ and $B$ sites inequivalent. The values of $n_A$ and $n_B$ are determined self-consistently. Although the problem has been simplified, it still poses a considerable challenge to solve. Selected results, which are obtained within the static mean-field approximation, have been recently presented by Y. Chen \textit{et al.} in Ref. \onlinecite{Chen_mf_cavity} and by N. Dogra \textit{et al.} in Ref. \onlinecite{Dogra_cavity}. In our paper we use the B-DMFT\cite{byczuk_bdmft} approximation to solve this problem. The previous studies showed that this method is well suited for studying the Bose-Hubbard type models.\cite{hofstetter_2species_2009, kauch_lce, dmft_snoek,anders_dmft,anders_dmft_short} It has also been applied to a model of a finite system inside the optical cavity.\cite{hof_cavity, reza_theo} We expand this research to infinite homogeneous system and present a more detailed study.


\subsection{B-DMFT for a bipartite lattice}\label{sect2b}

In the B-DMFT the self-energy is approximated to be momentum independent.\cite{georges_rmp} This allows us to use a self-consistent scheme in which we obtain local quantities by solving an effective local (``impurity'') problem and use Dyson equations to close the set of equations. A detailed derivation for the case of bosons on homogeneous lattice can be found in Ref. \onlinecite{byczuk_bdmft, anders_dmft}. In our work we consider a bipartite lattice with lower translational symmetry and, therefore, need to modify this procedure. Let us first notice that we have two distinct types of sites, corresponding to sublattices $A$ and $B$, which require different, impurity mapping. Its derivation is analogous to the homogeneous case, but needs to be performed separately for different sublattices. The result in a form of the action in the Feynman path-integral representation is following
\begin{widetext}
\begin{equation}\label{loc_action}
\begin{aligned}
S_{A/B}^{loc}= & \int_0^{\beta} \d \tau b^{\ast}(\tau)\left[\partial_{\tau}-\mu \mp V(n_A-n_B) \right] b(\tau)
 +\frac{U}{2}\int_0^{\beta} \d \tau b^{\ast}(\tau)b^{\ast}(\tau)b(\tau)b(\tau)
 -\kappa\int_0^{\beta} \d \tau \mathbf{\Psi}^{\ast}_{A/B}\mathbf{b}(\tau)\\
& +\frac{1}{2}\int_0^{\beta} \d \tau \int_0^{\beta} \d \tau' \mathbf{b}^{\ast}(\tau)\mathbb{\Delta}_{A/B}(\tau-\tau')\mathbf{b}(\tau').
\end{aligned}
\end{equation}
\end{widetext}
Here we use subscripts $A$ and `$-$' sign if we consider an impurity on the sublattice $A$ and use subscripts $B$ and `$+$' sign if the impurity is on the sublattice $B$. We also use a notation in which $\beta=1/T$ is inverse of the temperature ($k_B=1$), $\kappa=zt$, the number of nearest neighbors on the square lattice is $z=4$, $\tau$ is the imaginary time and 
\begin{equation}
\mathbf{b}=\left(
\begin{array}{l}
b\\
b^{\ast}
\end{array}
 \right)
\end{equation}
are the complex variables in the Nambu notation.\cite{Negele} Finally, in \eqref{loc_action} appear two external fields, the vector $\mathbf{\Psi}_{A/B}$ and the matrix $\mathbb{\Delta}_{A/B}$, which also depend on the sublattice type, hence the subscript. These are given, in close analogy with the homogeneous case, by
\begin{equation}
\mathbb{\Delta}_{A/B}(\tau-\tau')=-\sum_{i,j\neq 0} t_{i0}t_{j0} \langle T_{\tau} \mathbf{\hat{b}}_i(\tau) \mathbf{\hat{b}}_j^{\dag}(\tau') \rangle^{(0)}_{A/B},
\end{equation}
and 
\begin{equation}\label{phi}
\mathbf{\Psi}_{A/B}=\langle \mathbf{\hat{b}}_i\rangle^{(0)}_{A/B},
\end{equation}
where $i$ is a nearest neighbors of site $0$ (impurity) and $\langle \ldots \rangle^{(0)}_{A/B}$ stands for the {\it connected} part of the equilibrium average in the grand canonical ensemble of the system with site $0$ removed (independently on which sublattice the impurity resides we always assign to it an index $0$). Notice, that depending on the sublattice on which the impurity resides these averages will be different, which is reflected by the subscript $A$ or $B$. The operator $\hat{\mathbf{b}}^\dag = \left( \hat{b}^\dag, \hat{b} \right)$ is the Nambu notation for the creation and annihilation operators and $T_\tau$ represents time ordering of the operators. To summarize, for the impurity on one sublattice the physical quantities depend on: (i) a local potential $\mu$ due to the external reservoir, (ii) a local interaction, (iii) an effective local potential $-V(n_A - n_B)$ due to infinite-range interaction, (iv) an effective coupling to the surrounding sites (from another sublattice) represented by two types of fields, $\mathbf{\Psi}_{A/B}$ describing coupling to the condensate and $\mathbb{\Delta}_{A/B}$ describing coupling to normal particles.

Solving the impurity problem is computationally the most demanding step in the B-DMFT self-consistency loop. In this paper we use the continuous-time quantum Monte-Carlo (CT-QMC)\cite{gull_ctqmc,prokofev_qmc} as a single impurity solver.\cite{anders_dmft, anders_dmft_short} It is a stochastic method, which does not impose any extra approximations. Within this approach one can, in principle, obtain arbitrary accuracy of the results with the main limitation coming from the computation time. Most importantly we obtain following local quantities: $n_{A/B}$ average local occupation (sublattice dependent), $\phi_{A/B}=\langle \hat{b}_{A/B} \rangle$ the order parameter on impurity on sublattice $A$ or $B$, respectively, and $\mathbb{G}^{imp}_{A/B}(\i\omega_n)$, the impurity Green function on sublattice $A$ or $B$ in Matsubara frequencies. The latter can be used in the local Dyson equation in order to obtain the local self-energy
\begin{equation}\label{loc_Dys}
\mathbb{\Sigma}_{A/B} (\i\omega_n) = \i\omega_n \bbsigma_3 + \mu \mathbb{1} -\mathbb{\Delta}_{A/B}(\i\omega_n) - \left(\mathbb{G}_{A/B}^{imp}(\i\omega_n)\right)^{-1},
\end{equation}
where $\bbsigma_3$ is the Pauli matrix with $1$ and $-1$ on the diagonal.

As we already mentioned, in the B-DMFT the self-energy is approximated to be purely local. A direct consequence of this is that knowing the local part we have the full knowledge of the self-energy, within the approximation. This means that one can use a full lattice Dyson equation in order to obtain the updated Green function for the entire lattice $\mathbb{G}_{ij}(\i\omega_n)$. The full equations are presented in App. \ref{AppA}.

Finally, the last step in the B-DMFT procedure is to calculate new, updated values of the fields $\mathbb{\Delta}_{A/B}$ and $\mathbf{\Psi}_{A/B}$. For the former quantity we use again the local Dyson equation \eqref{loc_Dys}, however, instead of the impurity Green function we use its updated local value $\mathbb{G}_{A/B}$ obtaining
\begin{equation}\label{loc_Dys2}
\mathbb{\Delta}_{A/B} (\i\omega_n) = \i\omega_n \bbsigma_3 + \mu \mathbb{1} -\mathbb{\Sigma}_{A/B}(\i\omega_n) - \left(\mathbb{G}_{A/B}(\i\omega_n)\right)^{-1}.
\end{equation}
For the latter quantity we note, that the site of sublattice $A$ ($B$) is surrounded by sites of sublattice $B$ ($A$) and, therefore, couples to the condensate amplitude $\phi_B$ ($\phi_A)$. We also note, that the average in \eqref{phi} is over lattice with a cavity, and this change in the geometry of the system has to be taken into account. The resulting formula, in analogy with its counterpart for a homogeneous system, Ref. \onlinecite{anders_dmft, kauch_lce, moja,eckstein_noneq_2014}, has the following form
\begin{equation}{
\Psi_{A/B} = \phi_{B/A} +\frac{1}{\kappa}\left(\Delta^{(11)}_{A/B} (0) + \Delta^{(12)}_{A/B} (0) \right)\phi_{A/B}.}
\end{equation}
Notice that some indices are inverted.


\subsection{Metastability and phase transition line}\label{sect2c}
A characteristic feature of a self-consistent iterative method is that the converged solution might depend on the initial condition, from which the iteration starts. This is not a problem for the phase transition between MI and SF phases of the model (\ref{hamiltonian})-- if there exists a SF solution it has a lower value of the grand potential than the MI solution and therefore represents the true phase.\cite{anders_dmft} However, this issue does influence other phase transitions which we study in this paper. This might be easily understood in the atomic limit (hopping amplitude $t$ is set to zero) for zero temperature and for $\mu=0.4U$, as an example. In such case it is possible to solve the original lattice problem (\ref{hamiltonian}). There are two states that are candidates for the ground state. The average value of the Hamiltonian has local minimum with respect to small variations from these two states. One state corresponds to the MI with average occupation $n_A=n_B =1$ and one corresponds to the DW with $n_{A}=2$ and $n_{B}=0$. The physical solution is the one with lower value of the grand potential, which in the zero temperature is $\langle \hat{H}\rangle$ (notice, that the chemical potential $\mu$ has been included in the Hamiltonian (\ref{hamiltonian})). Thus we obtain a phase transition between MI and DW at $V=0.5U$. However, if we consider a problem mapped onto a single impurity and treat it in a self-consistent manner\footnote{Notice, that for $t=0$ static mean-field and the B-DMFT are equivalent.} we get a DW phase stable down to $V=0.3U$. This is because for $0.5U>V>0.3U$ the DW is a metastable solution. Similarly, for $V>0.5$ it is possible for self-consistent steps to converge to a MI solution, even though it has a higher value of the grand potential than the DW solution.

Therefore, solving the B-DMFT equations self-consistently is not enough to determine the phase diagram. In order to determine the physically true phase for a given set of parameters one needs to compare values of the grand potential for all of the metastable solutions. Here we revert to an approximate scheme of calculating the grand potential $\Omega$ by assuming that $\Omega\approx\langle\hat{H}\rangle$. This approximation is equivalent to neglecting the entropic contribution $-TS$, which becomes formally rigorous only in the zero temperature limit. We checked within the static mean-field approximation\cite{fisher} that the neglected term is small compared to the internal energy, owing to the low temperatures in which we performed calculations ($TS\sim 10^{-3}\langle\hat{H}\rangle$). Using this approximation has a negligible influence on the results in the greater part of the phase diagram. However, the small $-TS$ term becomes significant in the vicinity of the SS-DW transition driven by the change of hopping amplitude. This issue will be discussed later in Sec. \ref{sect3a}.


\section{Results}\label{sect3}

\subsection{Phase diagram and static properties}\label{sect3a}
\begin{figure}[pt!]
\resizebox{1.0\columnwidth}{!}{
\includegraphics{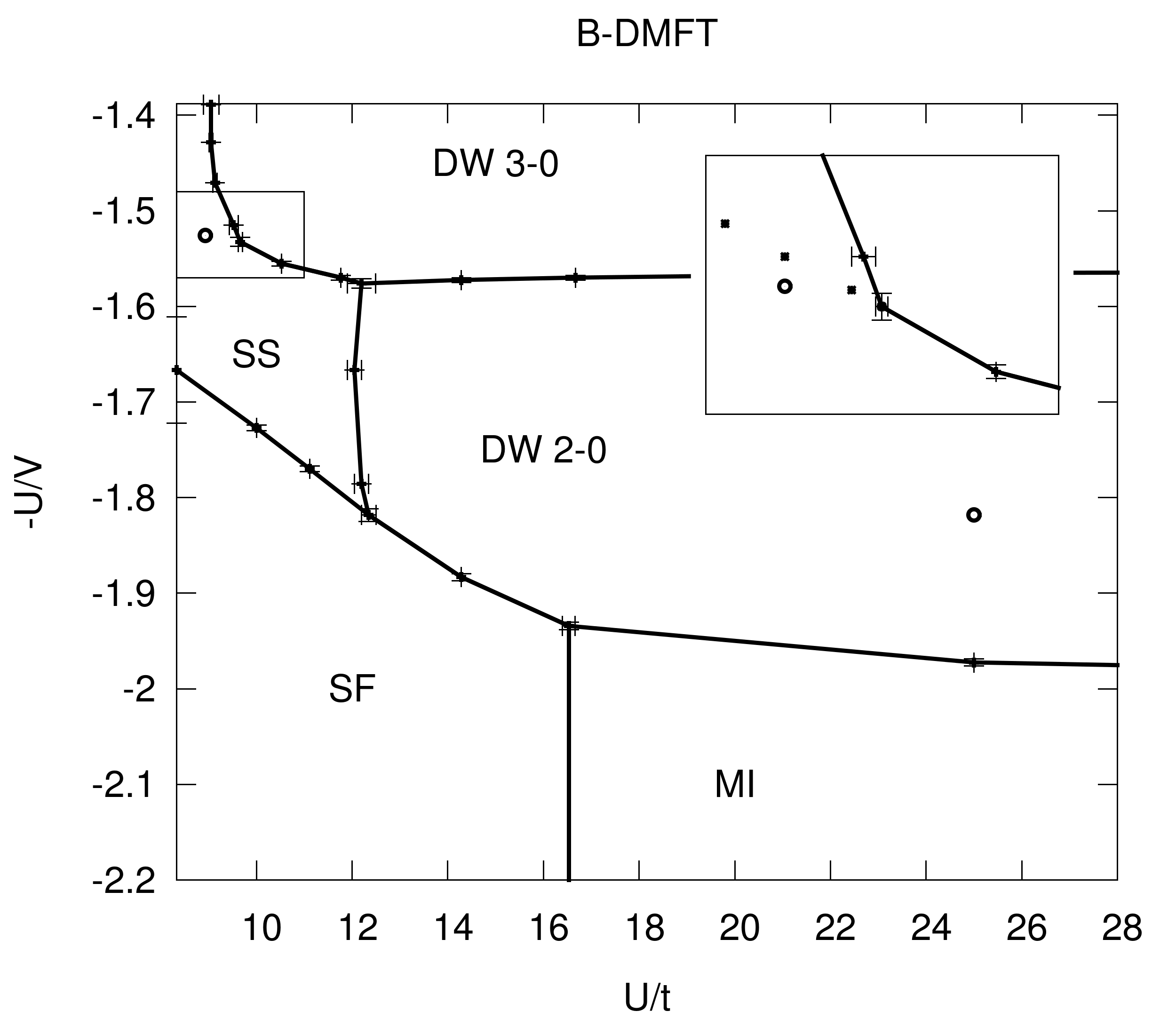}}
\resizebox{1.0\columnwidth}{!}{
\includegraphics{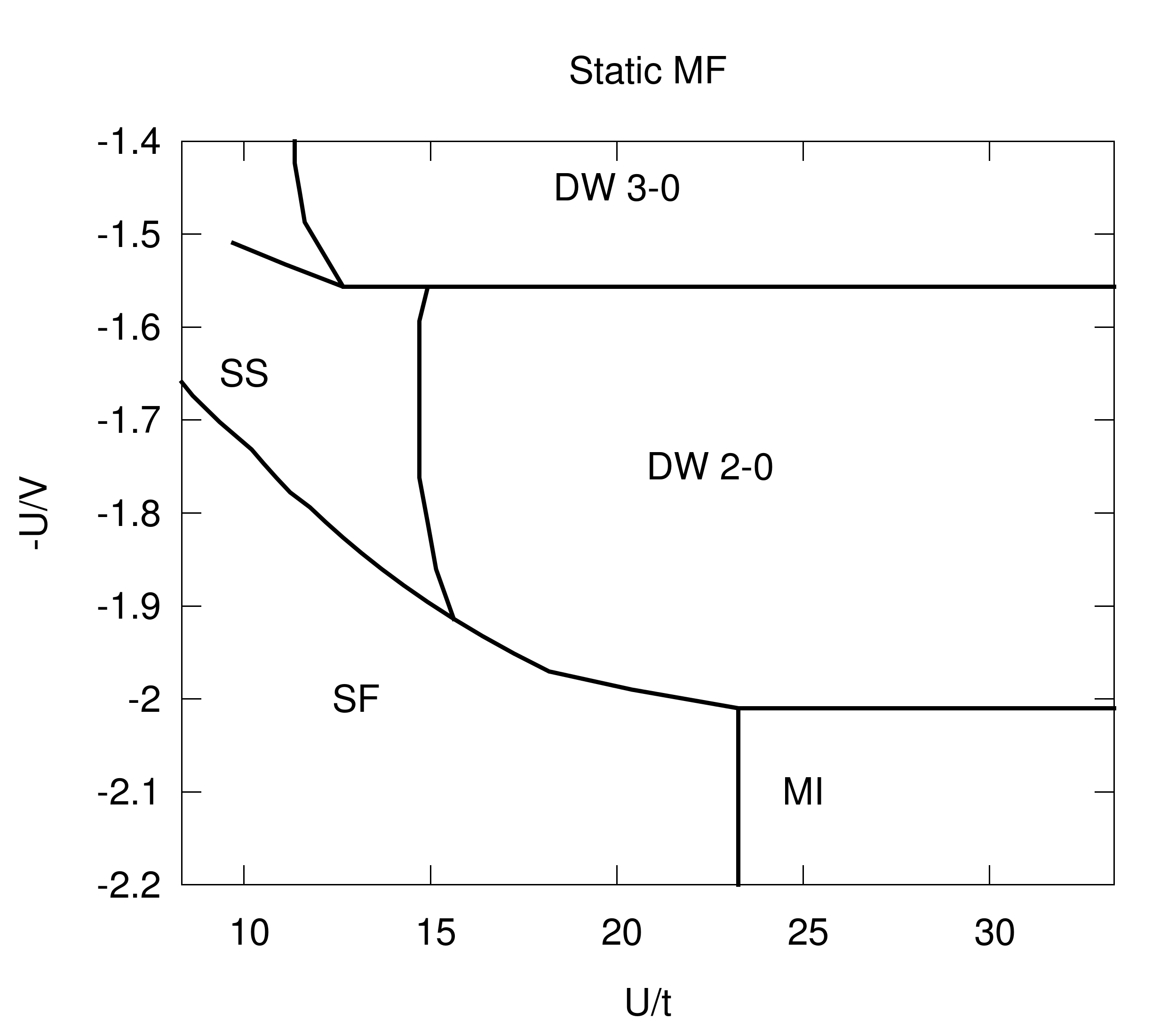}}
\caption{\label{phase_diag} Phase diagram of the two-dimensional system described by Hamiltonian (\ref{hamiltonian}). Parameters are set to $U=10$, $\mu=0.4U$ and $\beta=2$. Distinct phases are denoted: MI-- Mott insulator; SF-- superfluid; DW 2-0-- density wave with $n_A\approx 2$  and $n_B \approx 0$; DW 3-0-- density wave with $ n_A \approx 3$  and $n_B \approx 0$; SS-- supersolid. Top panel: results obtained within the B-DMFT method. The inset shows a close up of the area marked with rectangle. Empty circles denote parameters, for which the spectral functions are determined. Small dots in the inset denote kinks in the dependence of $\phi_A-\phi_B$ on $V$. Bottom panel: results obtained within the static mean-field method.
}
\end{figure}%
Our main goal is to expand the previous B-DMFT studies of a two-dimensional system in an optical cavity.\cite{hof_cavity, reza_theo} This is achieved by determining a phase diagram of a system in the thermodynamic limit in the $(t, V)$ space, comparing with the results of the static mean-field study\cite{Chen_mf_cavity} and by providing an analysis of the type of the phase transitions. These diagrams are presented in Fig. \ref{phase_diag}: the B-DMFT results -- top panel, and the static mean-field results -- bottom panel. The parameters for which we performed calculations are $U=10$, $\mu=4$, $\beta=2$. In order to make a distinction between different phases we define two order parameters: any of the $\phi_{A/B}=\langle \hat b_{A/B} \rangle$ fields (a situation, in which only one of them is (non-)zero is impossible), and $\Delta n=n_A-n_B $. These two order parameters allow us to define four phases: Mott insulating (MI), superfluid (SF), density wave (DW) and supersolid (SS) phases, as follows
\begin{itemize}
\item the Mott insulating phase is characterized by vanishing of both order parameters, i.e., $\phi_A=\phi_B=0$ and $\Delta n=0$. In this phase particles are immobile at $t=0$, localized on lattice sites and distributed uniformly in the system.
\item the superfluid phase is characterized by the presence of the condensed bosons in the system, where $\phi_{(A/B)} \neq 0$, and the uniform distribution of particles in it, i.e., $\Delta n=0$.
\item the density wave is defined by $\Delta n\neq 0$ and $\phi_A=\phi_B = 0$. There are no condensed bosons in the system, however, the symmetry between sublattices is spontaneously broken.
\item the supersolid phase is obtained when both order parameters are non-zero. There are two simultaneously broken symmetries, $\mathbb{Z}_2$ between the sublattices and $U(1)$ for the phase of the macroscopic wave function of the condensate.
\end{itemize}
Within the DW phase we find yet another two phases differing in the approximate value of $\Delta n$: DW 2-0 with $\Delta n\approx 2$ in which sublattice $A$ is on average occupied by approximately 2 particles per site, and DW 3-0 for which $\Delta n\approx 3$ and sublattice $A$ is on average occupied by approximately 3 particles per site. Sublattice $B$ is almost empty in both cases. In general we expect more DW type phases for different parameters of the system, $V$, $\mu$, etc. The phase transition between such phases is signaled by a discontinuity of $\Delta n$, here as a function of infinite-range interaction strength $V$. Apart from the difference in $\Delta n$ the two phases appearing in the diagram in Fig. \ref{phase_diag} are similar in their properties and symmetry.

A comparison between the results of the experiment\cite{Landig_cavity} and the different theoretical approaches\cite{Chen_mf_cavity,Dogra_cavity} shows certain similarities. We find the same type of phases in both approaches. The shapes of the diagrams are also similar. E.g., consider the phase transition line which separates MI from DW for small hopping amplitude and SF from SS for large. As we go along this line from large to small values of $t$ it descends and then flattens out. The SF extends to higher values of $V$ than the MI. This is a common feature of both experimental and theoretical results. A good agreement of experiment and theory was also shown in the results of Ref. \onlinecite{reza_theo, reza_exp}. We find only one significant discrepancy between the theories and the experiment. The interpretation of experimental results suggests that there exists a point in the phase-space in which all four phases meet, c.f., Fig. 3 of Ref. \onlinecite{Landig_cavity}. In our phase digram, and similarly in the phase diagrams obtained within the static mean-field approximation,\cite{Chen_mf_cavity,Dogra_cavity} such a point does not exist. The Mott insulating and supersolid phases are always separated by the density wave and superfluid phases. 
\begin{figure}[pt!]
\resizebox{1.0\columnwidth}{!}{
\includegraphics{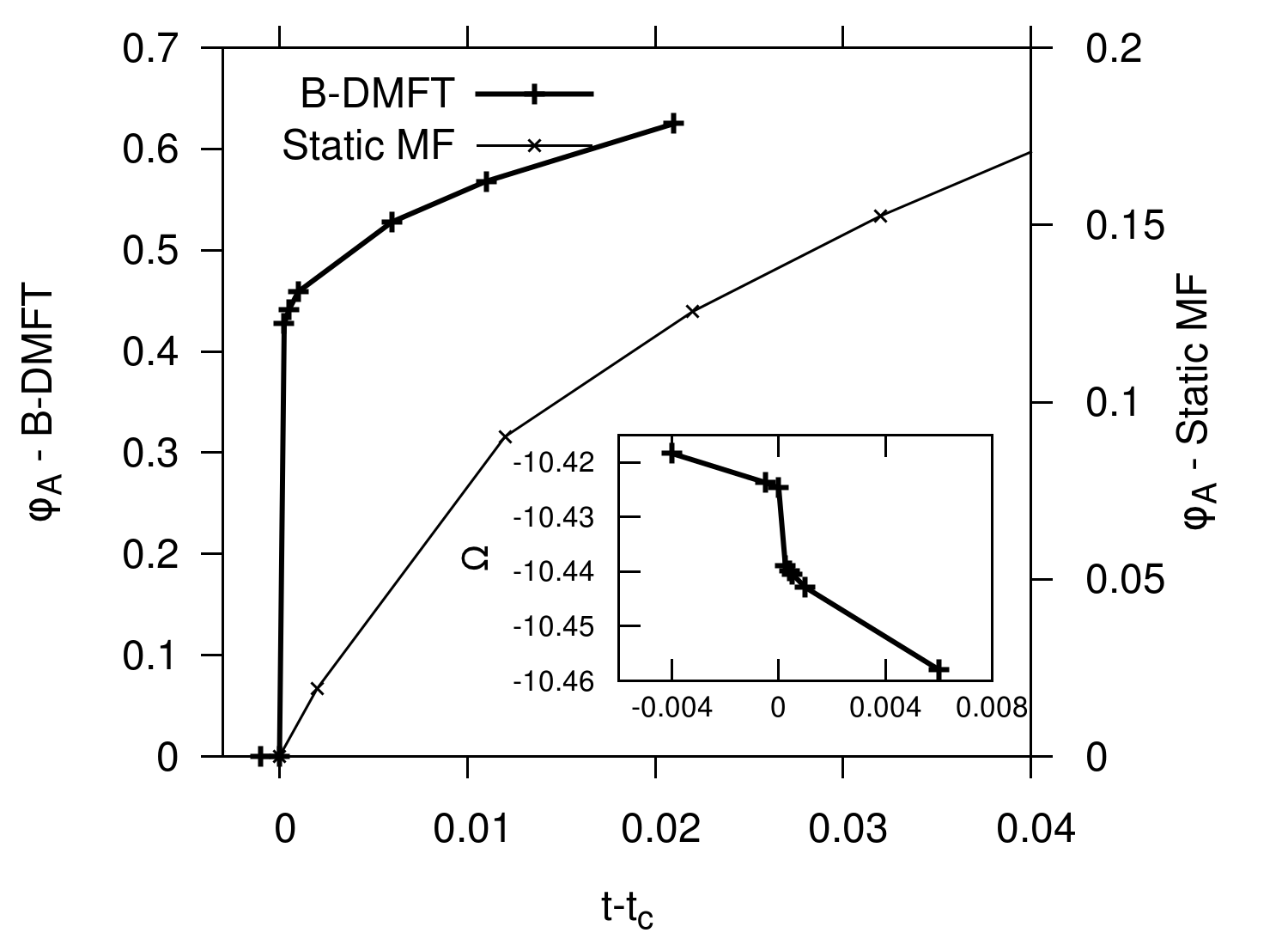}}
\caption{\label{ph_tr_ord} Dependence of the order parameter $\phi_A$ (of the doubly occupied site) on the relative hopping amplitude $t-t_c$ for $V=6$ at the phase transition between SS and DW. Thick line with `$+$' symbols represents results obtained with the B-DMFT method. Thin line with `$\times$' symbols represents results obtained with the static mean-field method. Apart from the different magnitude of $\phi_A$ obtained with the two methods, reflected by different scales on the graph, we observe that the behavior around critical point is significantly different. It seems, that $\phi$ is discontinuous in the B-DMFT, contrary to the static-mean-field results. Inset: dependence of the grand potential values on the hopping amplitude $t$. Its value for the SS phase is smaller and has a discontinuity as we cross the phase transition and the order parameter $\phi_A$ vanishes.}
\end{figure}%

We also find discrepancies between the B-DMFT and the static mean-field results. Firstly, there is a difference in the shape of the phase transition line between the DW and MI phases. In the bottom panel of Fig. \ref{phase_diag} we see, that this transition appears for a constant value of $-U/V\approx -2$. This is because the static mean-field is insensitive to changes of the hopping $t$ in the insulating phases. On the contrary, within the B-DMFT method the dependence on the hopping amplitude in the insulating phases is preserved. Hence, in the top panel of Fig. \ref{phase_diag}, the line separating DW and MI phases is not exactly flat but varies slightly with changing $t$.

The second discrepancy requires a more detailed study of the behavior of the order parameter in the vicinity of the phase transition lines. In the static mean-field study it has been observed that the type of the phase transition depends on the point at which we cross the phase boundary.\cite{Chen_mf_cavity,Dogra_cavity} E.g., for $U/t\approx 14.7$ and $-U/V\approx-1.67$ the phase transition between SS and DW 2-0 phases is continuous. At the same time, for $U/t\approx14$ and $-U/V\approx-1.56$ the phase transition between SS and DW 3-0 phases is discontinuous. Within our method this seems not to be the case. Every phase transition is discontinuous, except for the one between SF and MI, for which we have checked and confirmed previous results of Ref. \onlinecite{anders_dmft} (not shown here). This is particularly interesting for the transition from SS to DW 2-0 phase, because it shows a difference between the static mean-field and the B-DMFT results. The behavior of the order parameter close to the phase transition is depicted in Fig. \ref{ph_tr_ord}. In the static mean-field it is clearly continuous. On the contrary, in the B-DMFT it seems to drop abruptly at the critical value of the hopping amplitude, $t_c$. This conclusion is supported by the fact that no power law dependence $\phi_A\sim |t-t_c|^{a}$ fits to the data. We also observe that the grand potential is always smaller in the SS phase and has a discontinuity at the phase transition point, which is shown in the inset of Fig. \ref{ph_tr_ord}. The issue with the continuity can be attributed to the approximation which we use. The neglected entropic term is of the same order of magnitude as the jump in the grand potential. Taking it into account could heal the problem but should not introduce large changes in the phase diagram. The computation of the entropic contribution with sufficient numerical accuracy is, however, beyond our present implementation of the B-DMFT.
\begin{figure}[pt!]
\resizebox{0.95\columnwidth}{!}{
\includegraphics{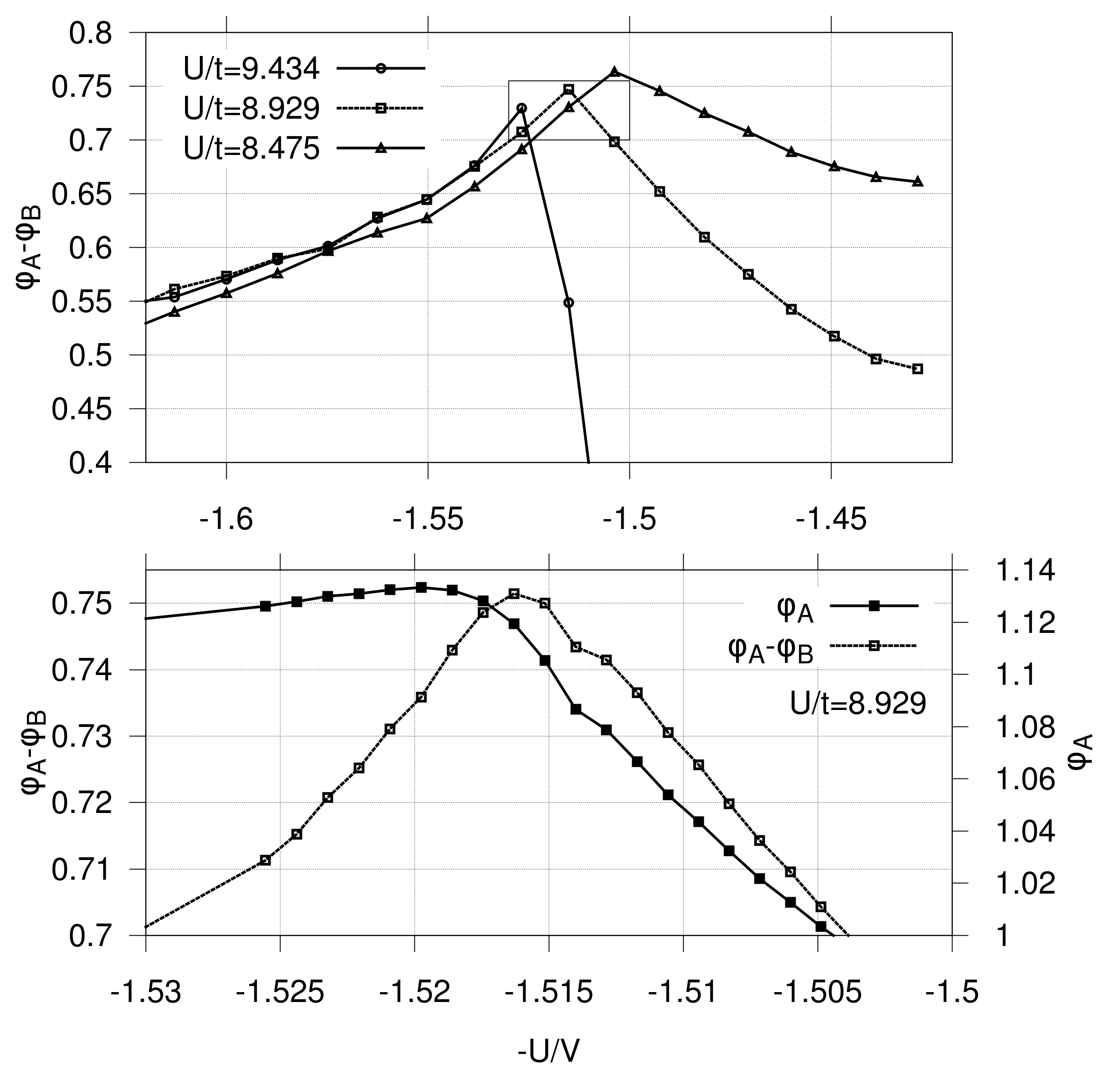}}
\caption{\label{kinks} Top panel: dependence of $\phi_A-\phi_B$ (difference of order parameters on sublattices $A$ and $B$) on infinite-range interaction term $V$ for several values of $U/t$, c.f., inset in Fig. \ref{phase_diag}, top panel. One can observe a sudden change of the slope of the function as $V$ increases. Bottom panel: A more detailed plot of $\phi_A-\phi_B$ for $U/t=8.929$ in the area marked with rectangle in top panel and comparison to behavior of the SF order parameter on sublattice $A$.}
\end{figure}%

It is also interesting to investigate the behavior of the order parameters within the SS phase. Particularly intriguing is the dependence of $\phi_A-\phi_B$ (difference of the condensate amplitudes on the sublattices) on the infinite-range interaction strength $V$ in the region presented in Fig. \ref{phase_diag}, inset of the top panel. The quantity $\phi_A-\phi_B$ as a function of $V$ is plotted in Fig. \ref{kinks}. The behavior is non-monotonic and seems to have a sharp ``kink''. A more detailed study, with a finer grid, shows that neither the derivative of $\phi_A-\phi_B$ nor the other order parameters, e.g., $\phi_{A/B}$, are discontinuous, c.f., Fig. \ref{kinks}, bottom panel. We also do not detect any change in the symmetry of the solution. Therefore, we conclude that the observed behavior does not represent a true phase transition but merely a crossover between a SS with $\Delta n\approx 2$ and a SS with $\Delta n\approx 3$. It would be worth investigating whether the situation does not change in the zero temperature, however, this is not possible with our method. The presented behavior represents one more difference between the B-DMFT and the static mean-field results.\cite{Dogra_cavity} Namely, in the latter the SS is not a single phase but rather splits into two (or more) phases separated by a phase transition line ending in a critical point, c.f., Fig. \ref{phase_diag}, bottom panel. 

\subsection{Spectral functions}\label{sect3b}

In the following we present the spectral functions of the model (\ref{hamiltonian}). We skip the discussion of the problem in the MI and SF phases since it has been already thoroughly studied.\cite{moja,huber,sengupta,dupuis_prl_2009,kopietz_prl_2009,knap_1d_2010,knap_sf_2011,zaleski,strand} On the other hand, the spectral functions of the SS and DW phases have not been investigated in details yet. The only study we are aware of is the one within the static mean-field approximation and only for the lowest-energy excitations.\cite{Dogra_cavity} In order to elaborate on that subject further we consider two types of spectral functions:
\begin{itemize}
\item {\it local density of states} $A_{A/B}(\omega)=-\frac{1}{\pi}\mathrm{Im}[G_{ii}(\omega)]$, where $G_{ii}(\omega)$ is the local Green function; subscript $A$ or $B$ specifies the sublattice to which $i$ belongs,
\item {\it momentum resolved spectral function} $A_{\alpha}(\mathbf{k},\omega)=-\frac{1}{\pi}\mathrm{Im}[G^{\alpha}_\mathbf{k}(\omega)]$, where the Green function $G^{\alpha}_\mathbf{k}(\omega)$ is represented in the basis of operators $\hat{b}_{\mathbf{k};1}$ and $\hat{b}_{\mathbf{k},2}$ which diagonalizes the noninteracting Hamiltonian, hence the index $\alpha \in \{1,2\}$.
Notice that since the lattice has lower translational symmetry, the area of the Brillouin zone (BZ) is reduced by half.
\end{itemize}
Within the B-DMFT method we obtain Green functions on the imaginary axis $G(\i\omega_n)$. In order to determine the Green functions on the real axis $G(\omega)$ we need to perform analytic continuation. We use the \textit{maximum entropy}\cite{jarrell} method for the numerical analytic continuation since within the CT-QMC we obtain results for finite number of frequency points and with a stochastic noise. More details on the data preparation and obtaining spectral functions within the B-DMFT can be found in Ref. \onlinecite{moja}.

\begin{figure}[pt!]
\resizebox{1.0\columnwidth}{!}{
\includegraphics{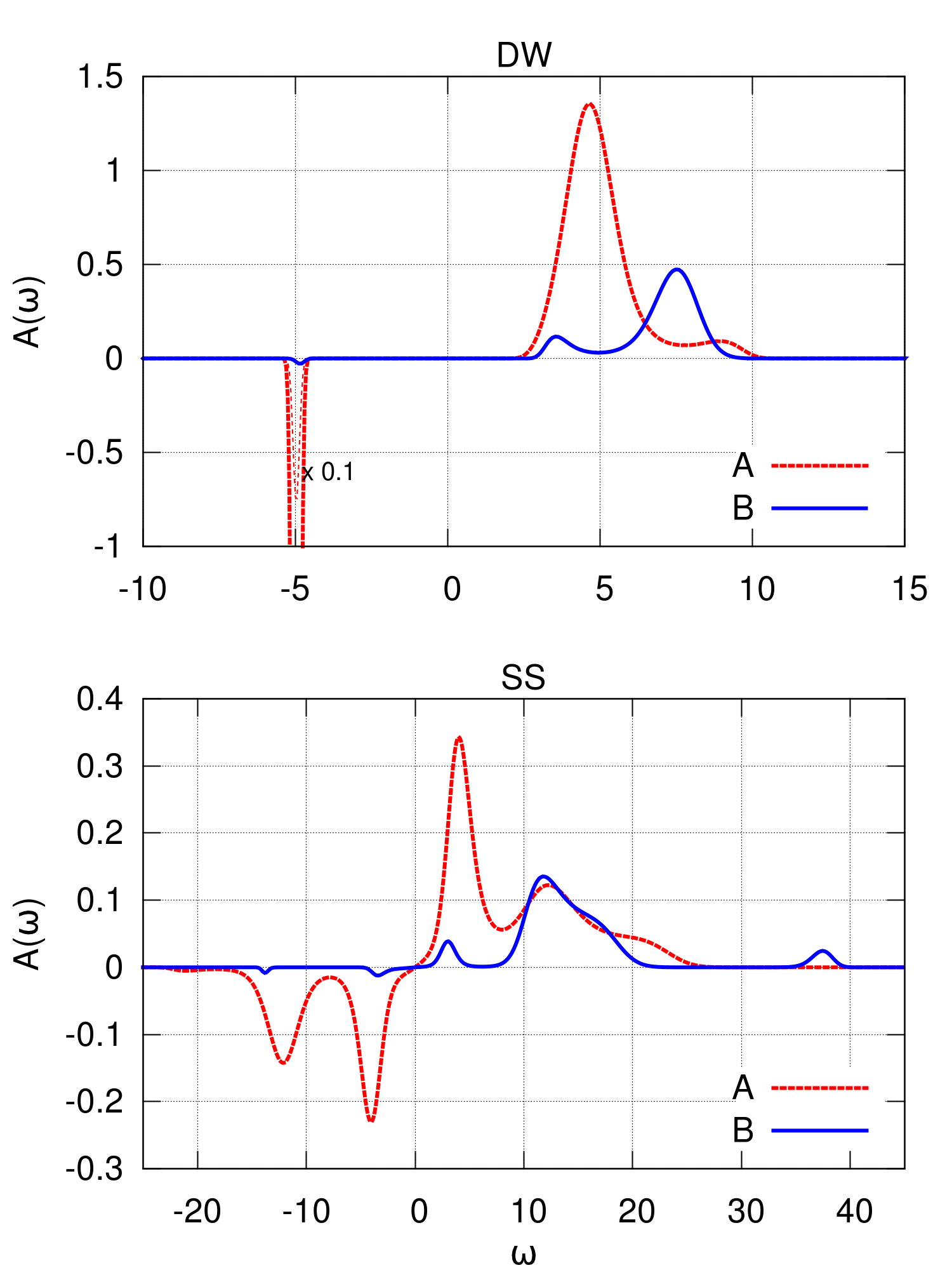}}
\caption{\label{spectr} Local densities of states $A(\omega)$ of the Bose-Hubbard model with infinite-range interaction. Parameters are set to $U=10$, $\beta=2$ and $\mu=4$. The values of $t$ and $V$ are: top panel -- $t=0.4$, $V=5.5$ which corresponds to DW phase; bottom panel-- $t=1.12$, $V=6.555$ which corresponds to SS phase. See Fig. \ref{phase_diag} for reference.}
\end{figure}

In Fig. \ref{spectr} we present results for the local density of states $A(\omega)$ for both sublattices, where $A$ is occupied and $B$ is nearly empty. The parameters were set to $U=10$, $\mu=4$ and $\beta=2$. We set $t=0.4$, $V=5.5$ to obtain the DW phase and $t=1.12$, $V=6.555$ to get the SS phase (see Fig. \ref{phase_diag} for reference). The results for the former are presented in the top panel. In this case the average occupation on different sites are $n_A= 1.9906$ and $n_B= 0.0096$, which means that one sublattice is almost doubly occupied at each site and the other is nearly empty. We distinguish three peaks, which we call bands on the basis of the momentum resolved spectral functions analysis, which will be discussed later. Two of those are particle bands and appear for positive values of $\omega$ and one is a narrow hole band appearing for negative values of $\omega$. The distance between the center of hole and the center of particle bands is approximately equal to the interaction strength $U=10$. The hole band centers at around $\omega=-5$. It has a form of a narrow peak and is present for both sublattices, however, its weight is significantly smaller for the sublattice $B$. This is because creation of a hole can occur only on the occupied site. Hence the peak is suppressed for the nearly empty sublattice $B$. The small width of the band comes from the fact that holes are localized on the subllatice $A$, bound to it due to absence of particles on sublattice $B$. This means that their dependence on quasi-momentum is weak. 

Let us consider the two particle bands. For the occupied sublattice $A$ more of the spectral weight is distributed to the band with lower energies, concentrating between $\omega =3$ and $\omega =6$. We also see a shoulder corresponding to the second band with higher energies. Conversely, on the empty sublattice we observe more of the spectral weight distributed to the higher energy band, spanning between $\omega=7$ and $\omega=9$ and a shoulder corresponding to the lower energy excitations. The fact that the bands are not completely separated can be attributed to the finite resolution of {\it maximum entropy} and/or to a finite temperature.

This behavior can be understood on a basis of the problem in atomic limit, that is with $t=0$. A similar comparative analysis was presented in Ref. \onlinecite{strand,strand2}. For given parameters and average occupations we have $V(n_A -n_B )\approx 11$. This would give two excitations on the occupied sublattice: hole excitation at $\omega=-5$ with $A(\omega)=-2\delta(\omega+5)$ and particle excitation at $\omega=5$ with $A(\omega)=3\delta(\omega-5)$. Similarly, on the empty sublattice this gives a particle excitation at $\omega=7$ with $A(\omega)=\delta(\omega-7)$. The nonzero value of $t$ results in broadening of the bands compared to the local problem, the lower band extends towards lower energies and the higher band, towards higher energies. $t \neq 0$ also results in some exchange of particles between sublattices hence the states become mixed and we see signatures of excitations corresponding to the $B$ sublattice on the sublattice $A$ and vice versa.

Next we consider the results for the SS phase, for which $ n_A\approx 2.45$ and $ n_B\approx 0.2$. They are presented in Fig. \ref{spectr} bottom panel. The bands are much wider because the hopping amplitude is larger than in the previous case. This is most prominent for the negative part of the spectrum of the sublattice $A$. In the DW phase we observed localized hole excitations, hence a narrow band. In the SS phase the interpretation of the part of spectrum with $\omega<0$ as hole excitations looses its virtue. This is due to the presence of the condensate: $\phi_A^2\approx1.27$ and $\phi_B^2\approx0.17$ and, as a result, fluctuating number of particles in the system $\langle \hat{b} \rangle\neq 0$. The elementary excitations combine the properties of both particles and holes, e.g., the Bogoliubov quasi-particle operator is a superposition of creation and annihilation operator.\cite{shi} Therefore, for each excitation with energy $\omega$ its spectral weight will be distributed between the peaks at $\omega$ and $-\omega$. The spectrum for the sublattice $B$ looks significantly different. The negative and low energy part of the spectrum appears due to presence of condensation-- fluctuations in its phase and amplitude. As condensate fraction is significantly smaller on this sublattice, these features in the spectral function are also much weaker. The main part of the spectrum starts at around $\omega=10$, which coincides with energy of adding a particle on an empty site of the sublattice $B$ with $V( n_A - n_B )\approx 14.4$, $\mu=4$ and for $t=0$ (atomic limit). Nevertheless, it is much wider than $zt$ (here $=4.48$), which seems to describe approximately the width of bands observed in the DW phase. The spectral weight is very small around $\omega=-10$ therefore the mixing of particle and hole properties of the excitations seems to be smaller for higher energy. Finally, we note that at high energy, $\omega\approx 37$, there appears yet another resonance, whose origin we cannot explain for now. As its weight is relatively small, its exact shape and position may not be reliably reproduced by the MaxEnt method of analytic continuation.
\begin{figure}[pt!]
\resizebox{1.0\columnwidth}{!}{
\includegraphics{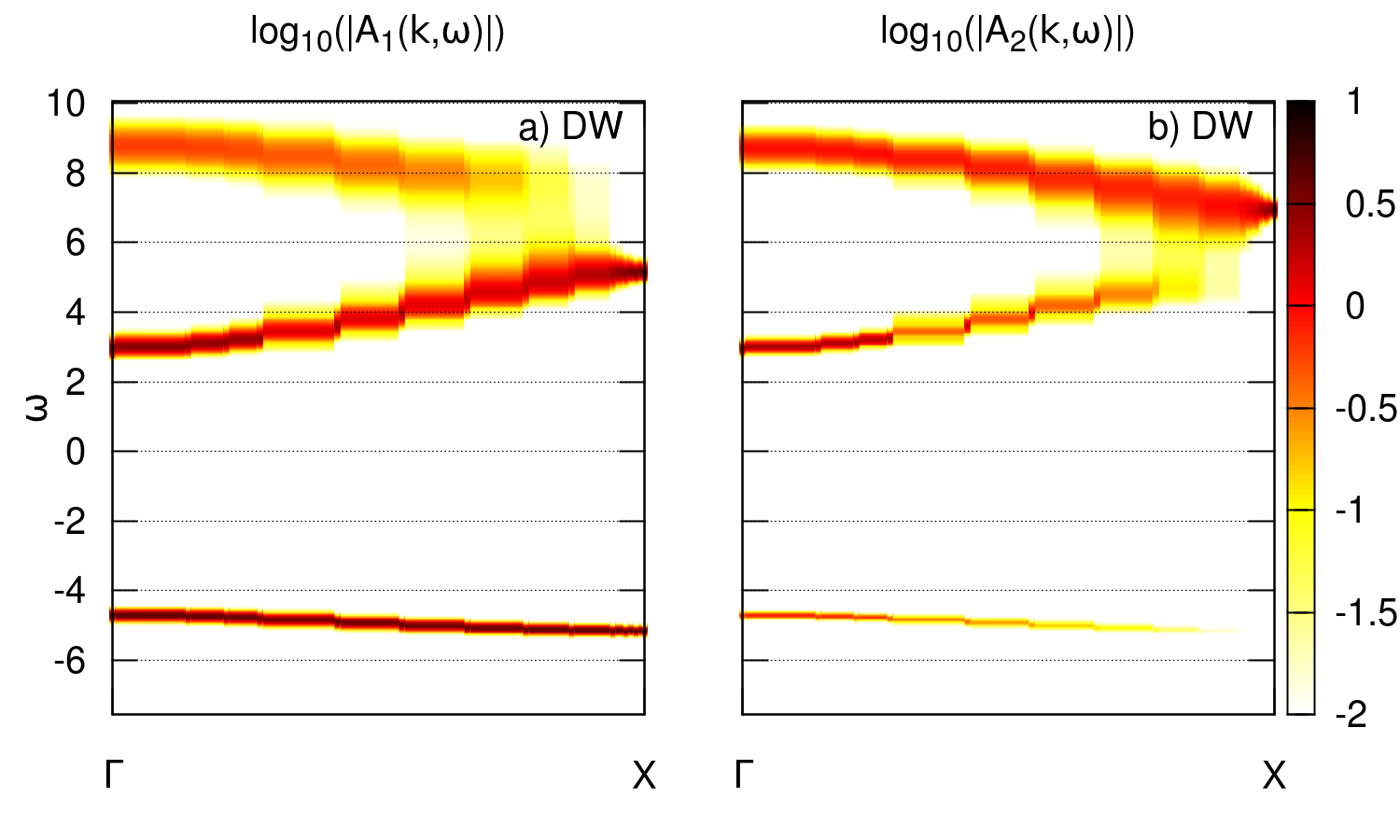}}
\resizebox{1.0\columnwidth}{!}{
\includegraphics{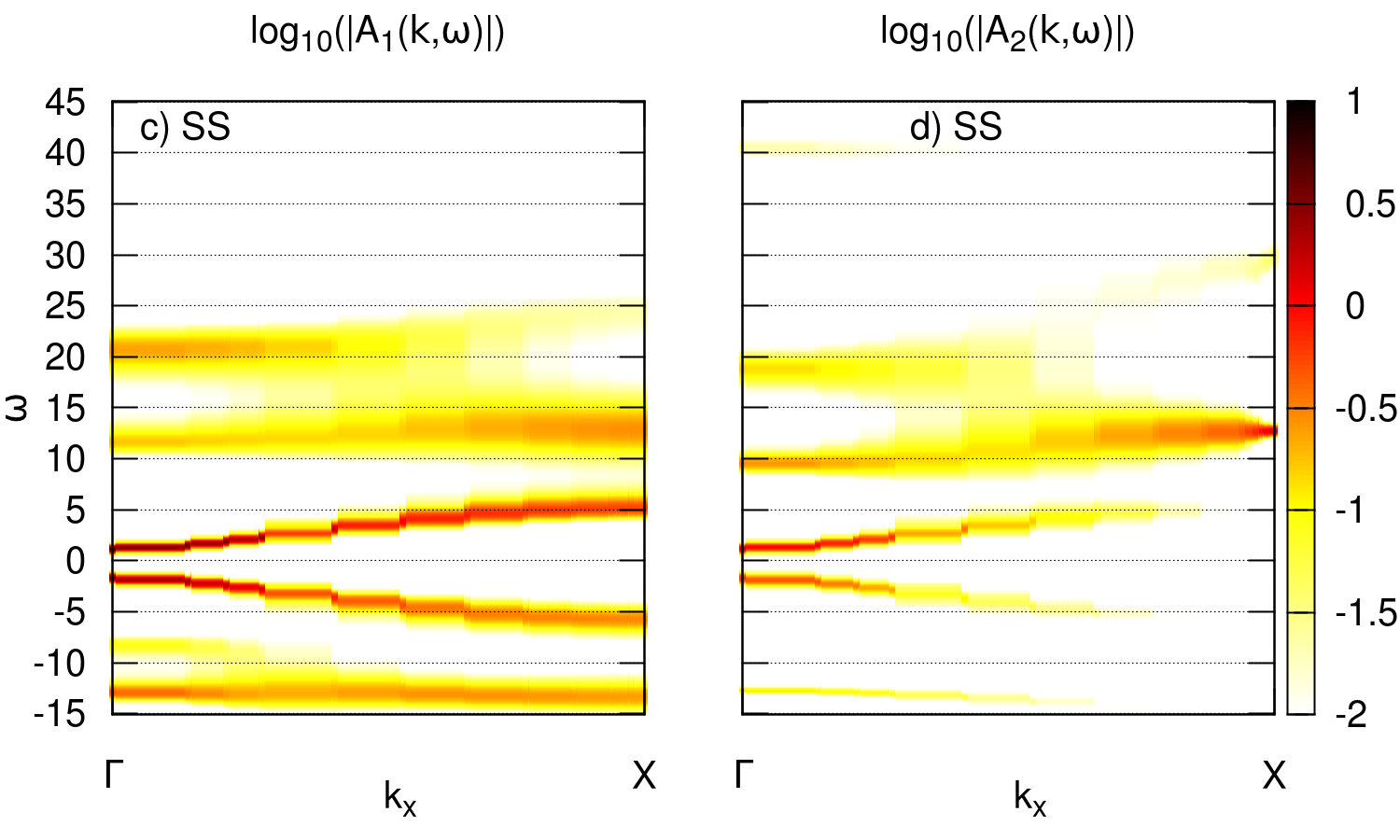}
}
\caption{\label{spectr_k} Momentum resolved spectral functions $A(\mathbf{k},\omega)$ of the Bose-Hubbard model with infinite-range interactions for the DW phase (top panels) and the SS phase (bottom panels). The parameters at which the calculations were performed are the same as in Fig. \ref{spectr}. Lower index of $A$ denotes one of two states withe quasi-momentum $k$, which diagonalize the non-interacting problem. Note the logarithmic color scale.}
\end{figure}%

In Fig. \ref{spectr_k} we present the momentum resolved spectral functions $A(\mathbf{k},\omega)$. Due to the lowering of the lattice translational symmetry the Brillouin zone (BZ) is reduced and, as a result, one needs two types of states for each value of the quasi-momentum. The operators for these states are chosen such that the Hamiltonian (\ref{hamiltonianMF}) without local interaction is diagonal in the new basis. We plot our results along the line between two special points, $\Gamma$ and $X$, in the reduced BZ (coordinates of these points in the original BZ are $\Gamma=(0,0)$ and $X=(\pi/2,\pi/2)$, assuming that lattice constant is equal to unity). It is important to note that dependence of $A(\mathbf{k},\omega)$ on $\mathbf{k}$ enters only through dispersion relation $\epsilon_\mathbf{k}$ of a noninteracting, homogeneous model, c.f., Eq. \ref{G_kq}. The spectral functions along different lines in the reduced BZ can be easily reproduced form values of $A(\mathbf{k},\omega)$ along the $\Gamma-X$ line.

As previously, we first analyze the results in the DW phase, shown in Fig. \ref{spectr_k} a) and b). The two plots correspond to two different operators for given quasi-momentum $\mathbf{k}$. The results are consistent with those of a local spectral function. We observe a narrow band (the dependence on $\mathbf{k}$ is weak) for the negative $\omega$. For the positive values of $\omega$ we observe two bands, one stretching from $\omega\approx 3$ to $\omega \approx 5$ and one from $\omega \approx 7$ to $\omega \approx 9$. This is in agreement with the results for the local spectral function, shown in Fig. \ref{spectr}. One should also notice that the gap between the two particle bands is approximately equal to $2(V (n_A-n_B) -U n_A)=2$. The same result would be obtained if we treated the interaction within the Hartree-Fock approximation. Therefore, in the DW phase, our approach simply reproduces the qualitative behavior, which is obtained within static mean-field consideration, with small quantitative corrections.

The plot of $A(\mathbf{k},\omega)$ looks significantly different in the SS phase, shown in Fig. \ref{spectr_k} c) and d). Firstly we observe two low energy bands. These bands seem to be symmetric with respect to $\omega=0$ axis. The energy of the excitations becomes small as we approach the $\Gamma$ point, almost reaching $\omega=0$. These low energy excitations originate on the sublattice $A$. The occupation on sublattice $A$ fluctuates between 2 and 3 particles ($n_A\approx 2.45$), therefore the energetic cost of adding or removing a particle on sublattice $A$ is expected to be low. However, due to the condensation the excitations are no longer particle or hole-like, but rather combine properties of both as explained earlier. In fact, judging by the homogeneous case we would expect these excitations to resemble Goldstone modes, fluctuations in phase of the order parameter. Formation of these quasi-particles also explains the symmetry between the bands. The open gap between the lowest energy bands (around $\omega=0$) is a feature of the B-DMFT approach. It appears because in the B-DMFT the Hugenholtz-Pines theorem\cite{hugenholtz} is not satisfied.\cite{anders_dmft,moja} In fact we would expect not only a closed gap, but also dispersion relation to become linear as we approach $\omega=0$.

We also observe higher energetic bands, for positive and negative $\omega$. As the different bands, both for $\omega<0$ and for $\omega>0$, have similar energy, it is hard to distinguish which excitation process they are related to. We attempted to identify the bands basing on the simple picture of a site in the atomic limit with additional coupling to a symmetry breaking field (c.f. static Fisher mean-field). The value of this field was taken from B-DMFT calculations and the resulting Hamiltonian was diagonalized obtaining its eigenvalues and through them the energies of the excitations (note that with a symmetry breaking field the occupation number states will no longer be eigenstates of the system). Details of this method can be found in Ref. \onlinecite{strand2}. This approach allows to interpret the nature of some of the bands. E.g., the bands with $\omega \approx -13,\ -8,\ 12$ seem to be in good agreement with such a modified atomic picture. However, the positions of other bands, with $\omega\approx 20,\ 30,\ 40$, are not captured properly. While for the last two this could be attributed to the resolution and accuracy of MaxEnt procedure, it cannot for the $\omega\approx 20$ band, as its weight is not small (c.f., Fig. \ref{spectr_k} c)). It rather seems that the dynamical corrections of the B-DMFT play an essential role here. Therefore, one should be cautious with using the atomic limit analogy to interpret the spectral features of the SS phase.

Another feature requiring better understanding is that while in the DW phase low-energy particle band has energy $\omega$ increasing with quasi-momentum $\mathbf{k}$ and the other particle band has energy decreasing with quasi-momentum, it seems not to be the case this in the SS phase. For the two lowest energy bands with $\omega>0$ the energy increases with quasi-momentum. It could be that a band with inversed dispersion appears for such energies that it overlaps with other bands making it hard to distinguish.

We conclude, that trying to find a simple intuition basing on atomic limit works well for the DW phase but not completely for the SS phase. The higher energy excitations are probably of a more complex nature-- dynamical processes are captured by the B-DMFT but not present in the the static picture. It would be interesting to investigate the momentum resolved spectral functions of the Bose-Hubbard model with infinite-range interactions with other methods, maybe even using stronger approximations, but providing results on the real axis and with better resolution.


\section{Summary}\label{sect4}

In summary, we have presented a thorough study of the Bose-Hubbard model with infinite-range interactions mediated by the cavity light modes. The use of the B-DMFT, which is a dynamical method, allows us to obtain a more reliable phase diagram. Because including the infinite-range interaction can lead to spontaneous breaking of the translational symmetry, we have derived and used an appropriate full self-consistency relation. The main result is the phase diagram. Comparison with other mean-field theoretical results shows both similarities and disagreements between the two approaches. While the phase diagram looks qualitatively similar, some phase transitions are of different type. We have also found an interesting behavior within the supersolid phase, which could be a precursor of a phase transition at zero temperature.

Apart from phase diagram we have studied the spectral properties of the supersolid and density wave phases. We have presented both local and momentum resolved spectral functions. We have analyzed our results by comparing with simple expansion around the atomic limit in small $t$ parameter, hoping to give some intuitive understanding of processes occurring in system with infinite-range interaction mediated by the cavity light mode.
\\

\begin{acknowledgments}
The authors would like to acknowledge fruitful discussions with J. Kune\v{s}, J. Skolimowski and D. Vollhardt and the constructive input of the First Referee. Support by the Deutsche Forschungsgemeinschaft through TRR 80 (K. B.) is acknowledged.
\end{acknowledgments}

\appendix

\section{B-DMFT self-consistency for a bipartite lattice with $A$ and $B$ sublattices inequivalent}\label{AppA}
In order to close the self-consistency of the B-DMFT we need to obtain the full Green function based on the local impurity results. Within the B-DMFT approximation one uses Dyson equation
\begin{widetext}
\begin{equation}\label{A1}
\mathbb{G}^{-1}_{ij} (\i\omega_n)=\\
\begin{pmatrix}
(\i \omega_n + \mu -V^{eff}_i-\Sigma_i^{11}(\i\omega_n))\delta_{i,j} - t_{ij} & -\Sigma_i^{12}(\i\omega_n)\delta_{i,j} \\
-\Sigma_i^{21}(\i\omega_n)\delta_{i,j} & (-\i \omega_n + \mu -V^{eff}_i-\Sigma_i^{22}(\i\omega_n))\delta_{i,j} - t_{ij}
\end{pmatrix},
\end{equation}
\end{widetext}
where we use similar notation as in Ref. [\onlinecite{moja}], Green function is in Nambu notation, $\Sigma(\i\omega_n)$ are matrix elements of the self-energy $\mathbb{\Sigma}$ (further on, for brevity, we do not write explicitly that $\Sigma$ depends on frequency), $V^{eff}=V( n_A- n_B)$ is the effective potential due to infinite-range interaction $V$, and $\omega_n$ are Matsubara frequencies. This general expression can be significantly simplified in the case of bipartite lattice with broken symmetry between sublattices in which case the self-energy and local potential are expressed as
\begin{equation}
\begin{split}
\mathbb{\Sigma}_{i}&=\bar{\mathbb{\Sigma}}\pm\delta\mathbb{\Sigma}, \\ V^{eff}_i &= \pm V^{eff},
\end{split}
\end{equation}
where we have `+' sign for sublattice $A$ and `-' sign for sublattice $B$, $\bar{\mathbb{\Sigma}}$ and  $\delta\mathbb{\Sigma}$ are halved sum and difference of self-energies on sublattices $A$ and $B$. Since the system is homogeneous we perform a Fourier transform. For convenience we choose the wave-vectors in the same way as for the system, where there is no difference between sublattices (notice, that this convention is different than the one used for spectral functions, see Sec. \ref{sect3b})
\begin{equation}\label{ft}
\mathbb{G}_{\mathbf{k},\mathbf{q}}^{-1}=\sum_{i,j}
\begin{pmatrix} \e^{\i \mathbf{k} \mathbf{R}_i} & 0 \\ 0 & \e^{\i \mathbf{k} \mathbf{R}_i} \end{pmatrix}
\mathbb{G}^{-1}_{ij}
\begin{pmatrix} \e^{-\i \mathbf{q} \mathbf{R}_i} & 0 \\ 0 & \e^{-\i \mathbf{q} \mathbf{R}_i} \end{pmatrix}/N.
\end{equation}
As a result we obtain the following expression
\begin{equation}
\begin{split}
&\mathbb{G}^{-1}_{\mathbf{kq}} (\i\omega_n)= \\
& (\i \omega_n \bbsigma_3 + (\mu-\epsilon_k)\mathbb{1} - \bar{\mathbb{\Sigma}})\delta_{\mathbf{k},\mathbf{q}} - (\delta\mathbb{\Sigma}+\mathbb{V}^{eff})\delta_{\mathbf{k},\mathbf{q}+\mathbf{\pi}}
\end{split}
\end{equation}
where $\epsilon_k=\sum_j t_{ij}\e^{\i k (R_i-R_j)}$ and $\mathbf{\pi}=(\pi,\pi)$ is a vector corresponding to the special point $M$ in Brillouin zone of a $2D$ lattice (lattice constant is set to unity). We notice, that the expression, apart from mixing pairs of states $\mathbf{k}$ and $\mathbf{k}-\mathbf{\pi}$, separates for different $k$'s. Inverting the above formula we obtain
\begin{widetext}
\begin{equation}\label{G_kq}
\begin{split}
\mathbb{G}_{\mathbf{kq}} (\i\omega_n)& = \left\{ \left[\bbsigma_3 \i \omega_n + (\mu-\epsilon_\mathbf{k})\mathbb{1} - \bar{\mathbb{\Sigma}}\right] - 
(\delta\mathbb{\Sigma}+\mathbb{V}^{eff})   \left[ \bbsigma_3 \i \omega_n + (\mu-\epsilon_{\mathbf{k}-\mathbf{\pi}})\mathbb{1} - \bar{\mathbb{\Sigma}} \right]^{-1}    (\delta\mathbb{\Sigma}+\mathbb{V}^{eff}) \right\}^{-1}
 \times  \\
& \left( \delta_{\mathbf{k},\mathbf{q}}\mathbb{1} + (\delta\mathbb{\Sigma}+\mathbb{V}^{eff}) \left[ \bbsigma_3 \i \omega_n + (\mu-\epsilon_{\mathbf{k}-\mathbf{\pi}})\mathbb{1} - \bar{\mathbb{\Sigma}} \right]^{-1}\delta_{\mathbf{k}-\mathbf{\pi},\mathbf{q}} \right).
\end{split}
\end{equation}
\end{widetext}
This expression can now be easily Fourier-transformed back, according to Eq. \ref{ft}, to the real-space, yielding lattice Green function, in particular its local part $\mathbb{G}_{ii}(\i\omega_n)$.


%

\end{document}